\documentclass[pre,amsmath,amssymb,floats,showpacs]{revtex4}

\usepackage{graphicx}

\def \beq{\begin{equation}}
\def \eeq{\end{equation}}
\def \bea{\begin{eqnarray}}
\def \eea{\end{eqnarray}}
\def \karman{$\mathrm{K\acute{a}rm\acute{a}n}$}

\begin{document}

\title{Spontaneous curvature cancellation in forced thin sheets}

\author{Tao Liang and Thomas A. Witten}
\affiliation{The James Franck Institute and the Department of Physics, The University of Chicago, 5640 S. Ellis Avenue, Chicago, IL 60637}

\date{\today}


\vskip 2cm

\begin{abstract}
In this paper we report numerically observed spontaneous vanishing of mean curvature on a developable cone made by pushing a thin elastic sheet into a circular container \cite{cerda-nature}. We show that this feature is independent of thickness of the sheet, the supporting radius and the amount of deflection. Several variants of developable cone are studied to examine the necessary conditions that lead to the vanishing of mean curvature. It is found that the presence of appropriate amount of radial stress is necessary. The developable cone geometry somehow produces the right amount of radial stress to induce just enough radial curvature to cancel the conical azimuthal curvature. In addition, the circular symmetry of supporting container edge plays an important role. With an elliptical supporting edge, the radial curvature overcompensates the azimuthal curvature near the minor axis and undercompensates near the major axis. Our numerical finding is verified by a crude experiment using a reflective plastic sheet. We expect this finding to have broad importance in describing the general geometrical properties of forced crumpling of thin sheets.  
\end{abstract}

\maketitle

\section{Introduction}
The crumpling of a thin sheet can be understood as the condensation of elastic energy into a network of two types of singular structures: folding ridges and point-like vertices. Scaling laws governing the energy and size of the ridge have been obtained analytically and tested numerically [2-5]. Point-like singularities are also studied extensively [1,6-12]. Besides the understanding of these singularities, however, there is a lack of knowledge about the geometrical and mechanical effects of confining forces on a thin elastic sheet. 

\begin{figure}[!htb]
\begin{center}
\includegraphics[width=0.8\textwidth]{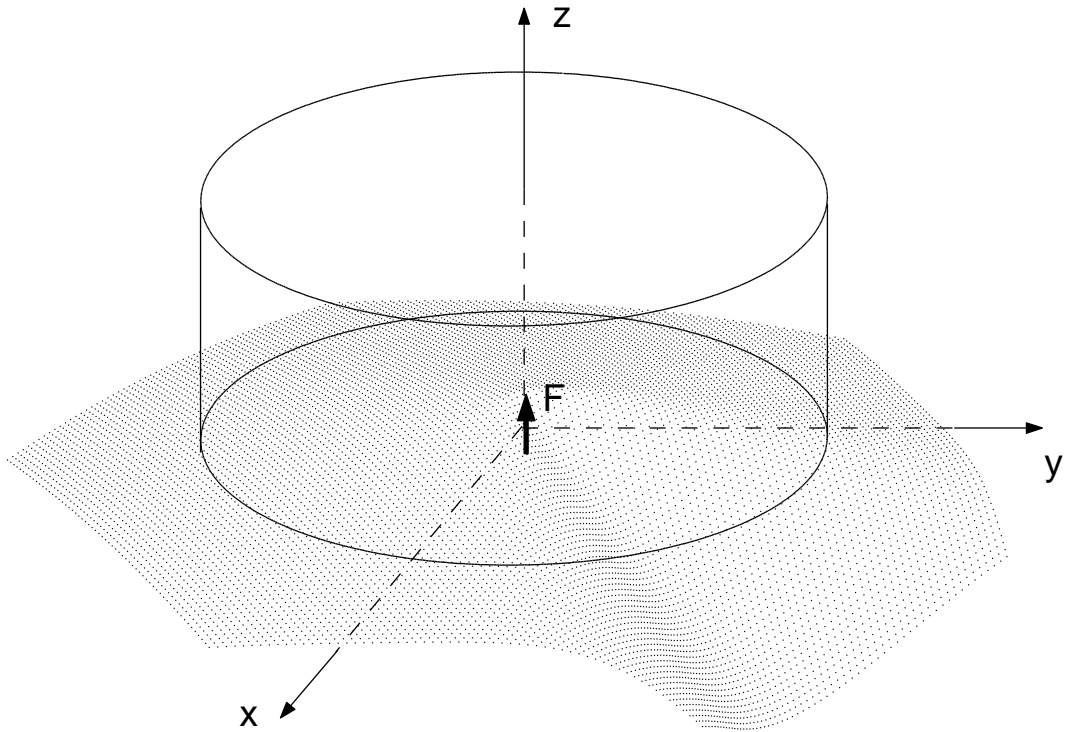}
\caption{A simulated developable cone, formed as we axially push a hexagonal elastic sheet into a cylindrical container. The force $F$ is applied to the central point of the sheet in the $z$ direction. The dots are the lattice points of our simulation, as described in Section II. The buckled region appears at the lower right region of the sheet, where it bends downward and away from the container.}
\label{shape}
\end{center}
\end{figure}

In this paper, we report an interesting observation of the response of an elastic sheet to the exertion of confining forces, using the developable cone or ``$d$-cone'' geometry studied by Cerda {\it et al} \cite{cerda-prl}\cite{cerda-nature}\cite{maha-new}. Specifically, this system is defined as follows. We push the center of a circular elastic sheet of radius $R_p$ axially into a cylindrical container of radius $R$ by a distance $d$. It is useful to express the deflection of the sheet by $\epsilon \equiv d/R$. Due to the constraint of unstretchability, some part of the sheet is buckled inwardly and the sheet deforms into a non-axisymmetric conical surface that is only in partial contact with the edge of the container (FIG.~\ref{shape}), thus forming a single developable cone. 

In the limit of pure bending, that is, as thickness goes to zero, the shape of $d$-cone has been shown to follow the solutions of the {\it Elastica} equation \cite{maha-new}. For asymptoticaly small deformation $\epsilon$, the shape can be analytically determined: this indicates that the buckled region occupies a fixed azimuthal angle of $138^\mathrm{o}$, independent of all relevant length scales \cite{cerda-prl}\cite{maha-new}. However, for a physical sheet with finite thickness $h$, as discussed in Ref. \cite{last}, stretching must occur and the stretching energy is concentrated in a small core region near the pushing tip. Outside this core region, bending energy strongly predominates over stretching energy. Cerda {\it et al} \cite{cerda-nature} propose that the size of the core region scales as $h^{1/3} R^{2/3}$. This scaling law was verified by our numerical simulations through direct geometrical measurements and through inferences from the measurements of central pushing force \cite{last}. However, our analytical estimates indicate that this scaling cannot by asymptotic. Away from the core region, the confining normal forces from the edge of the cylindrical container arise from the requirement to counteract the central pushing force. In the situation of pure bending, Cerda and Mahadevan \cite{maha-new} predict that the sheet exerts a concentrated point force to the container at the two take-off points where the sheet bends away from the container. Due to the translational symmetry along the region of the surface in contact with the container, the normal force is found to be independent of the azimuthal angle. 

The problem of interest here is the response of the sheet to these azimuthally constant normal forces from the edge of the container. Through our numerical study, we find that these normal forces produce a striking effect: they cause the mean curvature to vanish near the supporting edge within our numerical precision. In this work, we investigate this finding in a single $d$-cone as well as in variants of $d$-done chosen to explore the conditions for this striking phenomenon. The numerical method we used is specified in Section II, consistent with that of Ref. \cite{last}. We also describe our validation of its accuracy. In Section III, we give the detailed description of our numerical findings and effects of altering the systems in various ways. In Section IV, a simple experiment is performed to verify our numerical observations for $d$-cone. We conclude with Section V and discuss limitations of and implications from our findings. 

\section{Numerical Methods}
We begin by specifying the numerical models we use. An elastic sheet is modelled by a triangular lattice of springs of un-stretched length $a$ and spring constant $k$, after Seung and Nelson \cite{nelson}. Bending rigidity is introduced by assigning an energy of $J(1-\hat{n}_1 \cdot \hat{n}_2)$ to every pair of adjacent triangles with normals $\hat{n}_1$ and $\hat{n}_2$. When strains are small compared to unity and radii of curvature are large compared to the lattice spacing $a$, this model bends and stretches like an elastic sheet of thickness $h=a\sqrt{8J/k}$ made of isotropic, homogeneous material with bending modulus $\kappa = J\sqrt{3}/2$, Young's modulus $Y=2ka/h\sqrt{3}$ and Poisson's ratio $\nu=1/3$. Lattice spacing $a$ is set to be 1. The shape of the sheet in our simulation is a regular hexagon of side length $R_p$. The typical value of $R_p$ is $60a$. 

To obtain a single $d$-cone shape, we need to simulate the constraining container edge and pushing force. As shown in FIG.~\ref{shape}, the edge lies in the $x-y$ plane and is described by equation $x^2+y^2=R^2$. Pushing in the center of the sheet is accomplished by introducing a repulsive potential of the form $U_{\mathrm{force}}(z_1)=-Fz_1$, where $z_1$ is the $z$ coordinate of the lattice point in the center and $F$ is the magnitude of the pushing force. This force is applied in the positive $z$ direction. The constraining edge is implemented by a potential of the form $U_{\mathrm{edge}}=\sum C_p H(z_i)/([(\sqrt{x_i^2+y_i^2}-R)^2+z_i^2]^4+\xi^8)$, where $\xi$, $C_p$ are constants and the summation is over all lattice points with coordinates $(x_i, y_i, z_i)$. $H(z)$ is the unit step function smoothed over a lattice spacing, which makes certain that this potential only acts on the lattice points that have already moved into the container (those with $z_i > 0$). We choose the range $\xi$ of the potential to be one lattice spacing. Thus the force decays rapidly once the lattice points go away from the edge. A discussion concerning the choices of the edge potential form is carried out in Section V. 

The conjugate gradient algorithm \cite{recipe} is used to minimize the total elastic and potential energy of the system as a function of the coordinates of all lattice points. This lattice model behaves like a continuum material provided that the curvatures are everywhere much smaller than $1/a$. This limitation restricts the values of deflection $\epsilon$ of $d$-cone to be below 0.25. 

We determine the curvatures approximately from each triangle in the sheet. For this measurement, we take the curvature tensor to be constant across each triangle. We calculate it using the relative heights of the six vertices of the three triangles that share sides with the given triangle \cite{brian}. The six relative heights $w_i$ normal to the triangle surface are fit to a function of the form
\beq
w_i=b_1+b_2 u_i+b_3 v_i +b_4u_i^2 +b_5 u_i v_i+b_6 v_i^2, ~~i=1, \ldots, 6
\eeq
where \{$u_i,v_i,w_i$\} are coordinates of the vertices in a local coordinate system that has $w$ axis perpendicular to the surface of the given triangle. This choice of local coordinate system ensures that $b_2$ and $b_3$ are negligible so that curvature tensors can be determined only from the coefficients of quadratic terms. In practice, our numerical findings do show that the values of $b_2$ and $b_3$ are on the order of $10^{-2}$ or lower. Therefore, curvature tensors follow immediately from the identification $C_{uu}=2 \times b_4, C_{vv}=2 \times b_6, C_{uv}=b_5$. Mean curvature $C$ is defined as half of the trace of curvature tensor: $C=(C_{uu}+C_{vv})/2$.

This lattice model of elastic sheet has been used to study both the ridge and the point-like singularities in crumpled sheets [2-5]\cite{last}. The accuracy of it has been tested in various ways. Using this model, Lobkovsky \cite{alex2} numerically verified the ``virial theorem'' for ridges that bending energy is five times the stretching energy, for asymptotically thin sheets. In the study of $d$-cone, the ratio of the normal force from the $\delta$-function term to that from the angle-independent term is found to be 0.70 numerically \cite{last}, compared well with the theoretical prediction 0.69 \cite{maha-new}. Also, the azimuthal profiles of curvature are in reasonably good agreement with the theoretical prediction as deformation of $d$-cone goes to zero \cite{last}. In addition, we tested our program by setting the lattice at different initial states and letting the program look for minimized energy state. We found the lattice converges to the same state, with agreements of pushing force and energies better than one percent for fixed deflection.

\section{Conditions for vanishing mean curvature}
We consider the mean curvature of the sheet along a radial line that lies in the non-buckled region. Ideally, for a regular conical surface, the mean curvature only comes from the azimuthal component $C_{\theta\theta}$ and follows a $1/r$ decay, where $r$ is the distance to the tip. For a $d$-cone described here, however, due to the pushing of the normal forces from the edge of container, the sheet has to experience a small deformation near the region where it touches the edge. More specifically, these normal forces cause a small inward deflection of the sheet and hence induce both curvature and strain in the contacting region. Noticing that the curvature induced is in radial direction and has opposite sign to $C_{\theta\theta}$, we expect the mean curvature to be reduced near the edge.

\begin{figure}[!hbt]
\begin{center}
\includegraphics[width=0.6\textwidth]{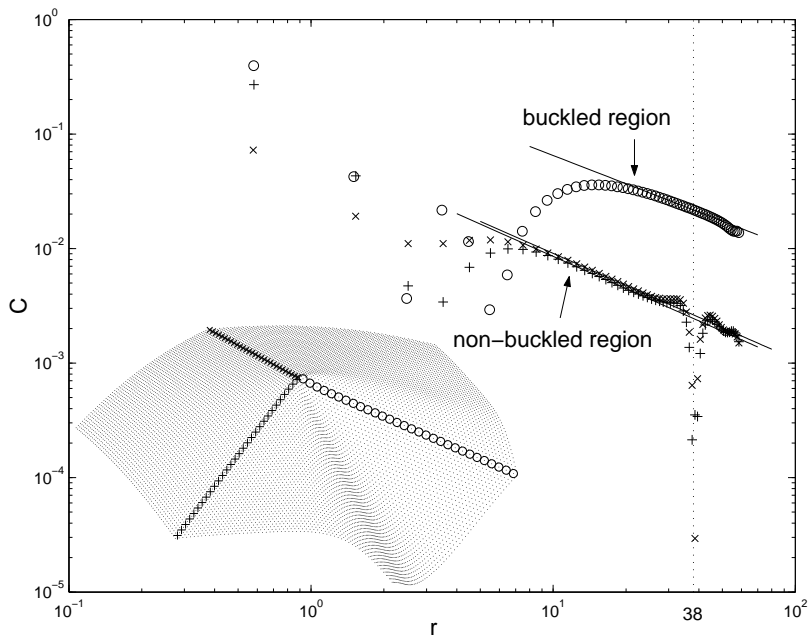}
\caption{Radial profiles of mean curvature in three different directions on a $d$-cone surface. The $d$-cone has thickness $h=0.102a$, confining radius $R=38a$ and deformation $\epsilon=0.15$, where $a$ is the lattice spacing in the numerical model \cite{nelson}. Inset shows the shape of the $d$-cone, where the three radial directions are denoted by the same symbols as the corresponding data. The slopes of fitted lines are $-0.934$, $-0.936$ and $-0.820$ for data denoted by pluses, crosses and circles, respectively.}
\label{curv_dist_all}
\end{center}
\end{figure}

Numerically, FIG.~\ref{curv_dist_all} shows the mean curvature profiles along three different radial directions on the same $d$-cone with thickness $h=0.102a$, deformation $\epsilon=0.15$ and cylinder radius $R=38a$. Data denoted by pluses and crosses are measured in the non-buckled region of the sheet; data denoted by circles are measured in the buckled region. Inset shows the $d$-cone shape and the three radial directions, denoted by the same symbols as the corresponding data. The values of three fitted slopes are all close to $-1$, verifying the $1/r$ feature of the mean curvature. However, for profiles in the non-buckled region, an abrupt drop of the mean curvature is observed near $r=38a$, where the sheet touches the edge of container. The curvature drops by more than one order of magnitude, making the mean curvature at the valley effectively zero. This indicates that the radial curvature caused by the pushing of normal forces from the edge not only reduces the mean curvature near the edge, but also virtually equals the original azimuthal curvature in magnitude.

\begin{figure}[!hbt]
\begin{center}
\includegraphics[width=0.55\textwidth]{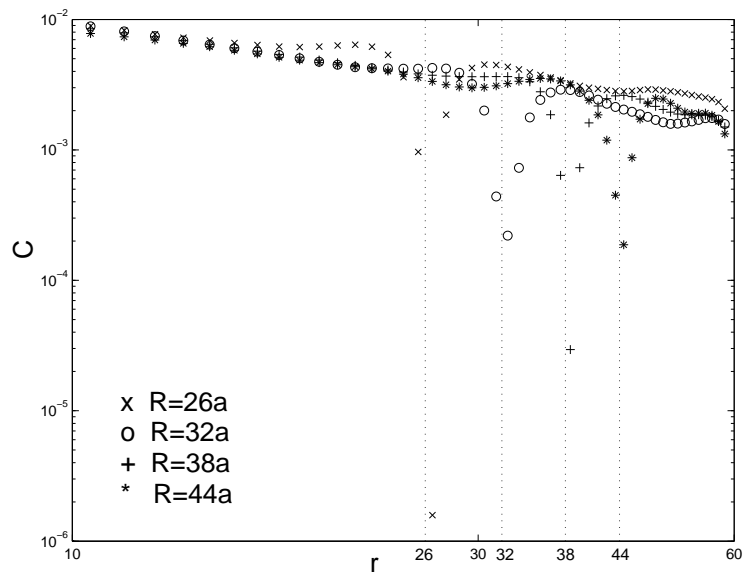}
\caption{Radial profiles of mean curvature for $d$-cones with four different values of confining radius: $R=26a,~32a,~38a,~44a$. Thickness $h=0.102a$ and deformation $\epsilon=0.15$.}
\label{var_r}
\end{center}
\end{figure}

\begin{figure}[!hbt]
\begin{center}
\includegraphics[width=0.55\textwidth]{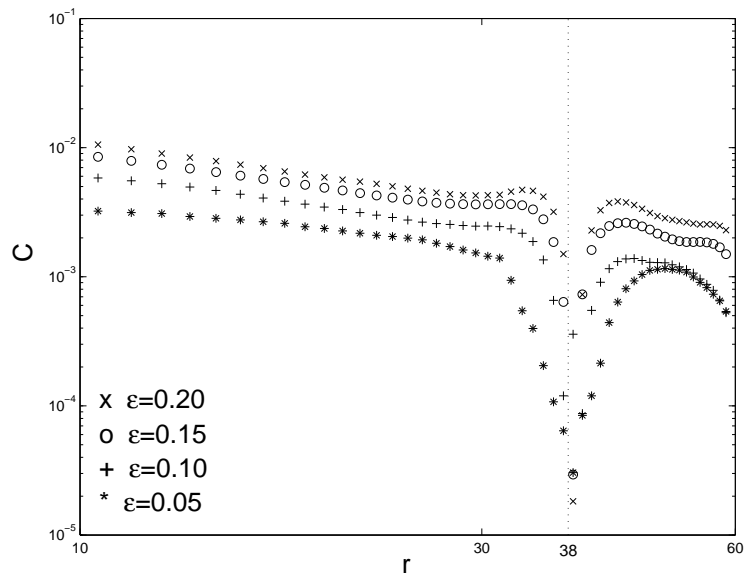}
\caption{Radial profiles of mean curvature for $d$-cones with four different values of deformations: $\epsilon=0.05,~0.10,~0.15,~0.20$. Thickness $h=0.102a$ and confining radius $R=38a$.}
\label{var_eps}
\end{center}
\end{figure}

We further observe that the feature of vanishing mean curvature does not depend on particular values of $R$ or $\epsilon$. FIG.~\ref{var_r} shows the radial profiles of mean curvature for $d$-cone formed with four different confining radius $R$, with fixed thickness $h=0.102a$ and deformation $\epsilon=0.15$. We observe that mean curvature drops down to zero effectively at each of the four radii. Moreover, FIG.~\ref{var_eps} displays the mean curvature profiles along a radial line on $d$-cone formed with four different deformations, with fixed thickness $h=0.102a$ and confining radius $R=38a$. It is seen that for four deformations, the mean curvatures all go to zero near $r=38a$, which the sheet touches the edge. 

To explore the conditions for the vanishing of mean curvature, we study variants of $d$-cone and make comparisons. To determine what is important in this phenomenon, we look at the the balance of normal force on each element of the sheet, which is expressed by the force von \karman~equation \cite{landau}
\beq
\partial_{\alpha} \partial_{\beta} M_{\alpha \beta} = \sigma_{\alpha \beta} C_{\alpha \beta}+ P~~.
\label{forcevon}
\eeq
here $M_{\alpha \beta}$ are torques per unit length, $\sigma_{\alpha \beta}$ are in-plane stresses, and $P$ is the external normal pressure on the sheet. For an asymptotically thin sheet, the azimuthal curvature of $d$-cone is $C_{\theta\theta}= \phi(\theta)/r$, where $\phi(\theta) = -\epsilon$ for $|\theta| > \theta_0$ (unbuckled region). In the limit of small deformation, it is found that $\phi(\theta) = -\epsilon \cos(a\theta)/\cos(a\theta_0)$ for $|\theta| \le \theta_0$ (buckled region), with $\theta_0=1.21$ rad and $a=3.8$ \cite{cerda-prl}\cite{maha-new}. The torques are proportional to curvatures through a constitutive law implicit in the energy equations \cite{mansfield}:$M_{rr}=\kappa \nu C_{\theta \theta} = \kappa \nu \phi/r$ and $M_{\theta \theta}= \kappa C_{\theta \theta} = \kappa \phi/r$, where $\kappa$ is the bending modulus and $\nu$ is the Poisson ratio. Using these laws to eliminate torques from the force von \karman~equation, one obtains
\beq
\sigma_{\theta\theta} = \frac{\kappa}{r^2} \left(2\nu + \frac{\ddot{\phi}}{\phi}\right) - \frac{Pr}{\phi}
\label{stress}
\eeq
where the dots above functions denote $\theta$ derivatives. For the unbuckled region, $\ddot{\phi}/\phi=0$, so $\sigma_{\theta\theta}=2\kappa\nu/r^2$. For the buckled region, $\ddot{\phi}/\phi = - a^2$, so $\sigma_{\theta\theta} = -\kappa (a^2-2\nu)/r^2$. Therefore, the azimuthal stress is compressive in the buckled region but tensile in the unbuckled region. Cerda and Mahadevan \cite{maha-new} show that for the same shape formed under a conical support, unlike a $d$-cone formed under a ring support as illustrated above, the azimuthal stress in the unbuckled region is compressive due to the exertion of normal pressure everywhere. In our case, the tensile azimuthal stress in the unbuckled region is changed to be compressive only near the edge of container, where the normal forces are nonzero. Whether the stresses are tensile or compressive, their magnitude goes as $1/r^2$. We call them type I stresses and denote them by symbol $\sigma^{(1)}$. It is noted that the type I stresses exist both in azimuthal and radial directions.    

Having looked at the local normal force balance of the surface, we now investigate the global force balance. We consider a region that encloses area between inner radius $R_c$ and outer radius $r$, where $r$ can take values between $R_c$ and $R$. $R_c$ denotes the size of the stretching core region. The tension exerted on the inner edge of this region is equivalent to the central pushing force of a regular $d$-cone. This force must be balanced by the force due to radial in-plane stress $\sigma_{rr}$ on the outer perimeter of the region. Let $\beta$ be the angle between a radial line or generator of the $d$-cone and the horizontal. Since $\tan{\beta}=\phi(\theta)$, we have $\sin{\beta} = \phi(\theta)/\sqrt{1+\phi^2 (\theta)}$. The balance of vertical forces yields
\beq
\int \sigma_{rr}  r \sin{\beta} d \theta = F ~~,
\label{above}
\eeq
which holds for every value of $r$ from $R_c$ to $R$. It is easy to see that the type I stresses alone can't satisfy this equation, since they go as $1/r^2$, they would give a $1/r$ prefactor on the left side of equation, while the right-side of equation is independent of $r$. Therefore, the integral from type I stresses must vanish and there must exist some additional tensile stresses in the outer region that scale as $1/r$ to satisfy Eq. (\ref{above}). We call these stresses as type II stresses and denote them by $\sigma^{(2)}$. They persist up to the supporting container edge, where normal force from the container counteracts the external pushing force. We can write $\sigma^{(2)} r = F e(\theta)$, where $e(\theta)$ is a function only of $\theta$ and satisfies $\int e(\theta) \sin{\beta} d \theta = 1$. It is obvious that $e(\theta)$ is of order unity. To estimate the magnitude of type II stresses, we notice that $F=\partial E/\partial d=(\partial E/ \partial \epsilon)/R \approx \kappa/R$. Thus $\sigma^{(2)} \approx F/r \approx \kappa/(rR)$. The type II stresses are comparable with type I stresses only near the container edge since the ratio $\sigma^{(1)} / \sigma^{(2)} \approx r/R$ for $R_c<r<R$.

Now that we see that the contacting normal force and the stresses seem to matter, we vary the system in ways that alter them. For this purpose, we study several geometries that we call the cut cone, regular cone, puckered cylinder and elliptical $d$-cone, and compare them with regular $d$-cone. 

\begin{figure}[!hbt]
\begin{center}
\includegraphics[width=0.6\textwidth]{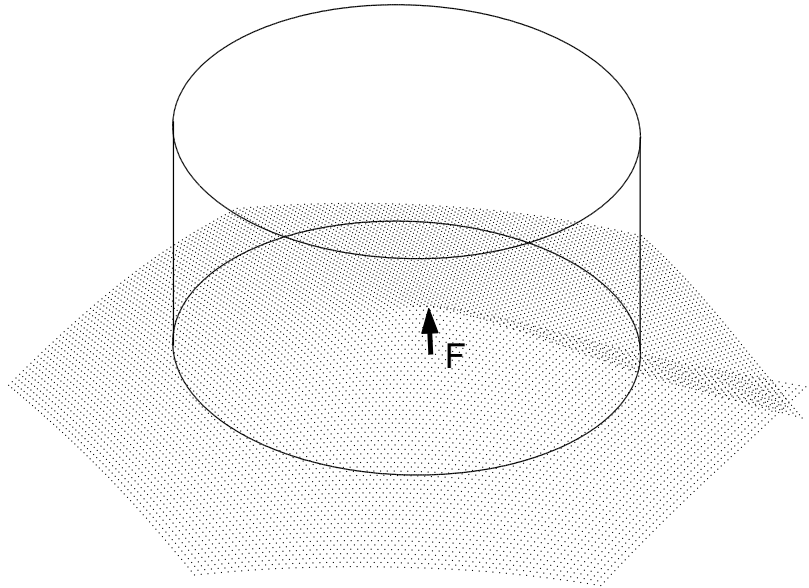}
\caption{After we cut the sheet along a radial line, the buckled region no longer appears as we push the sheet into the same container. Instead, the free boundaries overlap to accommodate the confining edge.}
\label{shape_cutcone}
\end{center}
\end{figure}

First, to study the effect of stresses, we cut the sheet along a radial line and push it into the same container to form a ``cut cone''. Now that the boundaries along the cut line are free to move, the buckled region no longer appears and the sheet will necessarily overlap itself in order to fit into the container, as illustrated in FIG.~\ref{shape_cutcone}. The cut cone mostly follows a conical shape as the unbuckled region of $d$-cone does, indicating that the azimuthal tensile stress remains the same as $d$-cone. Near the free boudaries, however, the sheet deviates from a conical shape and does not contact against the confining edge. Instead, there is a section of the sheet near the free boundaries that has less curvature than the corresponding $d$-cone shape, due to the lack of bending moment there. Consequently, the free boundaries contact the container at a nonzero angle and the azimuthal stresses arise from the pushing force from the edge. Due to the decreases in the energies, the pushing force is reduced by a finite factor relative to that of $d$-cone. This causes the radial tensile stress to be reduced by the same factor. Numerically, it is found that the pushing force is reduced by about one order of magnitude. The measurements of the mean curvature along radial lines show rather small decreases in the contacting region, far from enough to make mean curvature vanish. These contrasts to the $d$-cone show that the appropriate amount of radial stress in the $d$-cone may play an important role for the mean curvature to vanish.

\begin{figure}[!hbt]
\begin{center}
\includegraphics[width=0.7\textwidth]{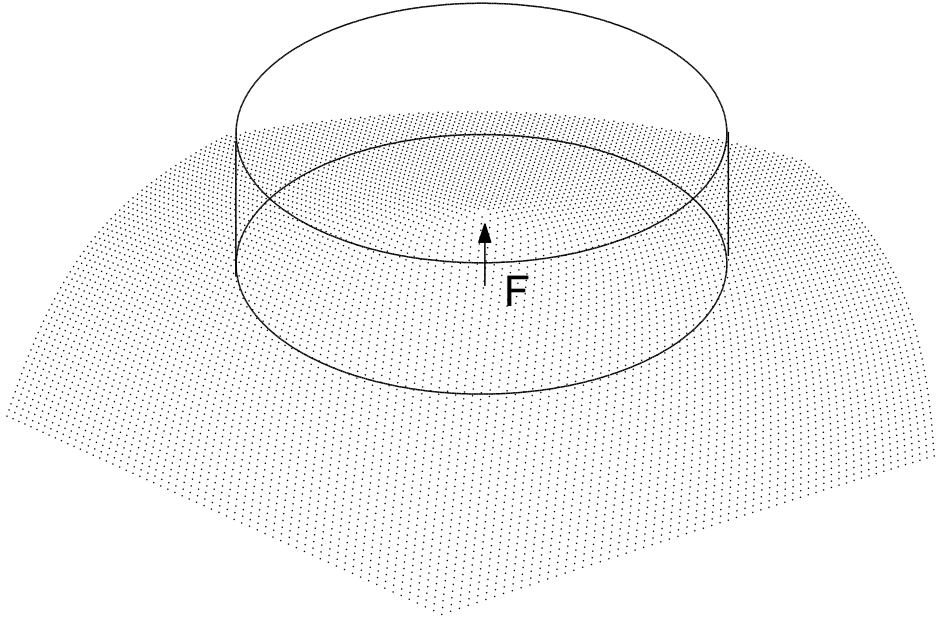}
\caption{A simulated regular cone formed by removing one sector and sewing the free boundaries.}
\label{shape_realcone}
\end{center}
\end{figure}

\begin{figure}[!hbt]
\begin{center}
\includegraphics[width=0.7\textwidth]{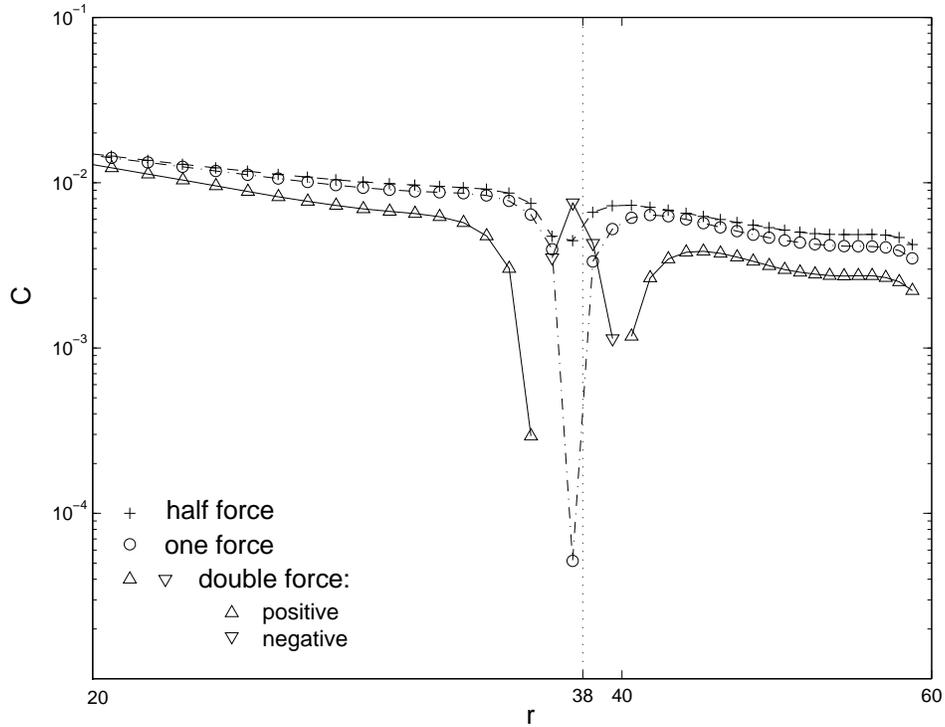}
\caption{Radial profiles of mean curvature of a regular cone pushed with three different forces against the confining edge. The regular cone has thickness $h=0.127a$ and deformation $\epsilon=0.66$. Confining radius is $R=38a$. Plus is for regular cone pushed with half $d$-cone force; circle is for regular cone pushed with one $d$-cone force; triangle is for regular cone pushed with double $d$-cone force, with downward triangles denoting mean curvatures changing sign.}
\label{threeforce}
\end{center}
\end{figure}

Next, we make a regular cone by removing one sixty-degree sector of the hexagon and joining the two free sides (see FIG.~\ref{shape_realcone}). We then push this regular cone into the same container as before, and measure the corresponding radial profile of mean curvature. The cone does not buckle for the range of forces we consider. We find that as we push with the similar force as that required for a $d$-cone of the same thickness and deformation, we observe the similar feature of vanishing mean curvature near $r=R$. To illustrate this point, FIG.~\ref{threeforce} shows the radial profiles of mean curvature of a regular cone pushed with three different forces against the confining edge. The pluses, circles and triangles are for a regular cone pushed by half of the corresponding $d$-cone force, one $d$-cone force and twice the $d$-cone force, respectively. We observe that the mean curvature vanishes as we push with one $d$-cone force, while it is not vanishing for half $d$-cone force and is changing sign for double $d$-cone force. When the central pushing force is similar, the normal forces from the edge are similar for regular cone and $d$-cone. So are the azimuthal compressive stresses induced by them in the contacting region. These facts show that the pushing force required to make $d$-cone is somehow just enough to produce the right amount of stresses for which the induced radial curvature cancels original azimuthal curvature. 

\begin{figure}[!hbt]
\begin{center}
\includegraphics[width=0.65\textwidth]{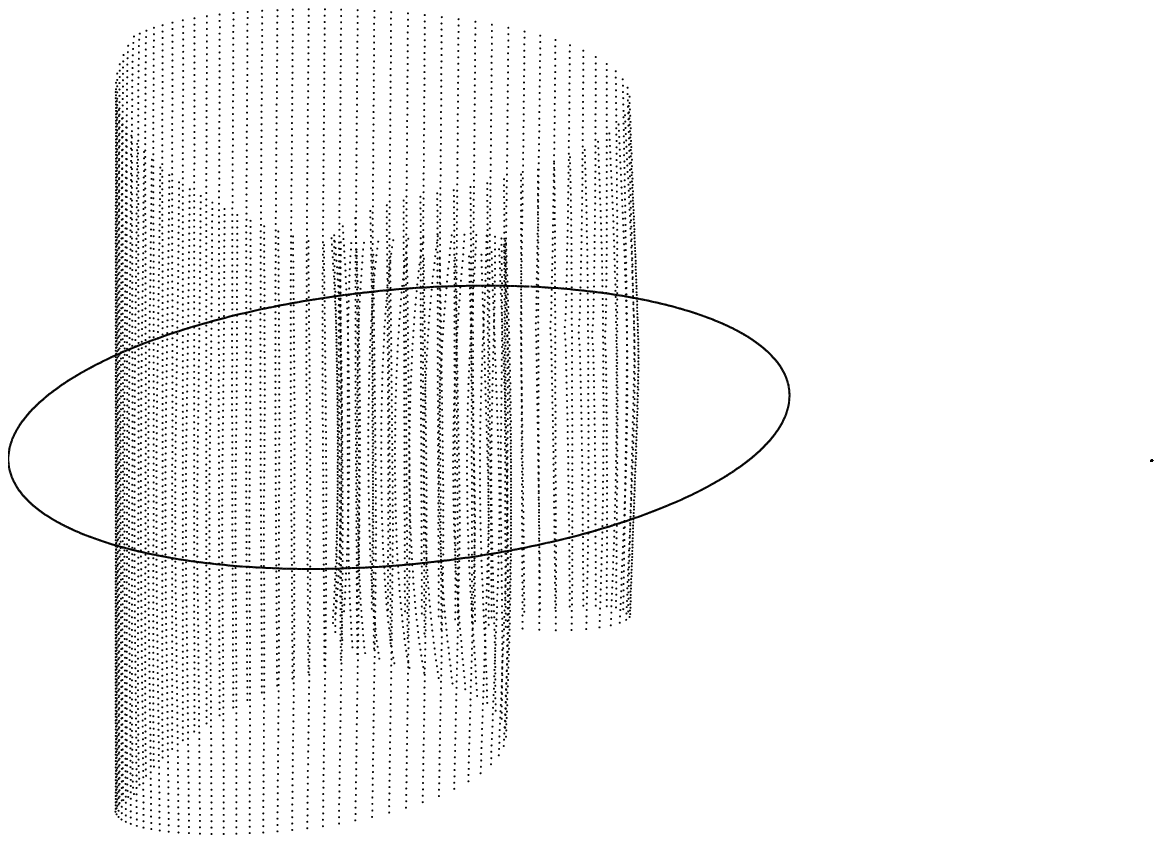}
\caption{A simulated puckered cylinder confined by a circular ring.}
\label{puckered_shape}
\end{center}
\end{figure}

We next investigate a variant that preserves the feature of a buckled sheet caused by a constraining boundary without the $d$-cone's feature of an additional force applied away from this boundary. To see this, we consider a puckered cylinder formed by rolling a sheet of cylindrical shape with radius $R$ into a circular cylinder of radius $b < R$, as studied by Cerda and Mahadevan \cite{maha-new}. To make closer comparison with the $d$-cone, we replace the cylindrical confining surface with a circular ring of the same radius. Now that the puckered cylindrical sheet is confined by the ring and the ring exerts normal forces on the sheet, we expect the mean curvature to be reduced in the contacting region where the sheet touches the confining ring. The simulated shape is shown in FIG.~\ref{puckered_shape}. We measure the mean curvature along a longitudinal line parallel to the axis of cylinder. A small drop of mean curvature is observed where the sheet touches the ring. Not surprisingly, the reduction of mean curvature goes up as the ratio $R/b$ increases. However, we find that the drop is no more than 10\% for various values of $R$ and $b$ that can produce reasonable confined shape, even for $R/b=2$. The reason that the mean curvature doesn't vanish near the contacting region is attributed to the absence of stess in the non-azimuthal direction. By contrast, the $d$-cone has a radial tensile stress which transmits the central force to the container. 

\begin{figure}[!hbt]
\begin{center}
\includegraphics[width=0.55\textwidth]{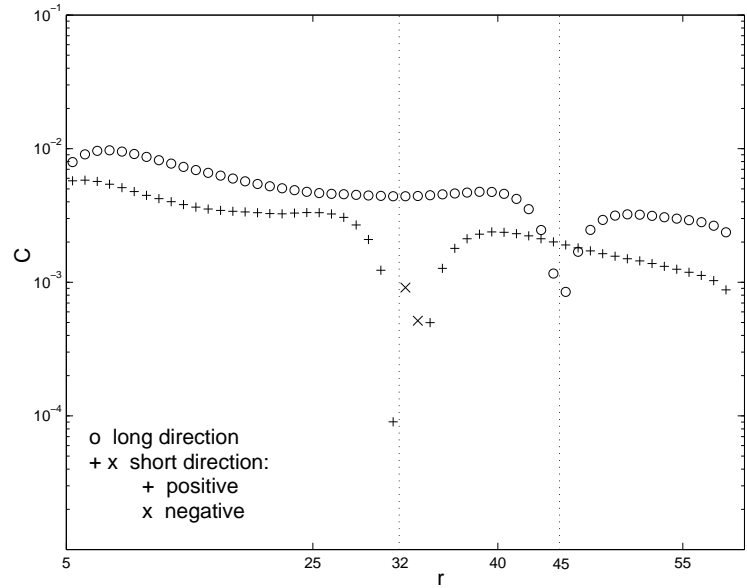}
\caption{Radial profiles of mean curvature in two different directions for the same $d$-cone confined by a elliptical edge with major axis of $45a$ and minor axis of $30a$. The distance is in linear scale. The data denoted by circles are measured along a major axis of the elliptical edge (long direction). The data denoted by pluses and crosses are measured in the direction that makes 60 degrees to the major axis (short direction). The crosses denote that the mean curvature becomes negative, so that it is overcompensated.}
\label{ellipse}
\end{center}
\end{figure}

Lastly, we observe that the circular symmetry is important for cancellation of mean curvature. To see this, an elliptical confining edge is used instead of a circular edge. Our original hexagonal sheet is pushed against this elliptical edge. The lengths of the major and minor axes of the ellipse are $45a$ and $30a$, respectively, with $a$ being the lattice spacing in the numerical model. FIG.~\ref{ellipse} shows mean curvature profiles in two different radial directions, both of which lie within the unbuckled region of the sheet. One profile is measured along a major axis of the elliptical edge. The other profile is measured along the lattice direction sixty degrees from the first and thus meets the ellipse close to the minor axis. For the sake of uniform comparison, we measured curvatures only along principal lattice directions. It is seen that the first profile shows a drop of mean curvature near the contacting region, although this drop is not as pronounced as those in FIG.~\ref{curv_dist_all}. On the other hand, the second profile shows that the mean curvature changes sign. The induced radial curvature is greater than the azimuthal curvature in magnitude, thus overcompensating the azimuthal curvature. We conclude that the break of circular symmetry in the unbuckled region leads to nonuniform degrees of compensations of mean curvature in different directions. 

\section{An Experimental Verification}
\begin{figure}[!hbt]
\begin{center}
\includegraphics[width=0.7\textwidth, bb=0pt 0pt 237pt 179pt]{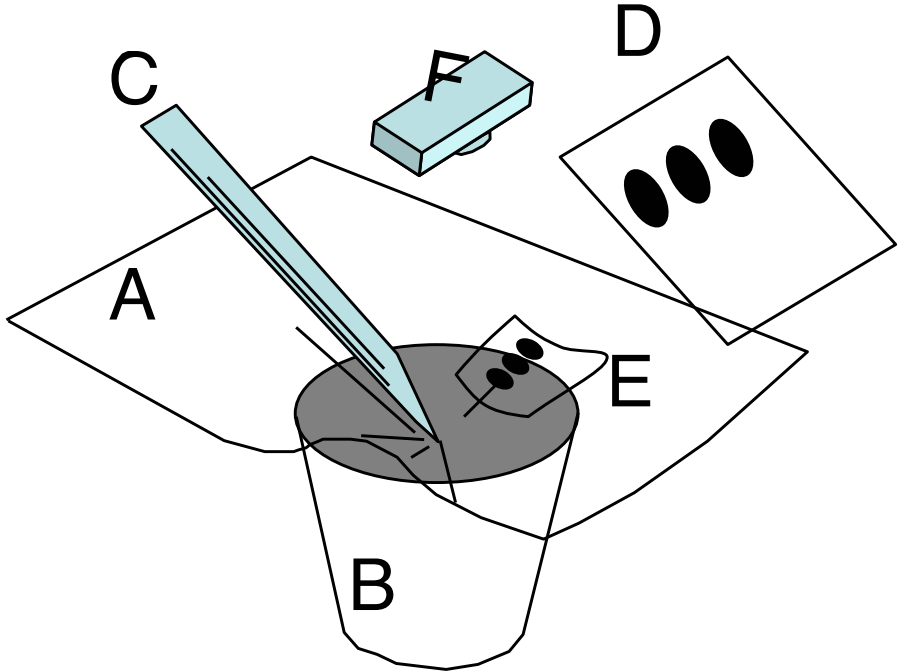}
\caption{(Color online) Sketch of apparatus for measuring curvature. Reflective transparency A is pressed into the container B by a probe C. Object D, a row of identical black circles on another transparency, forms the reflected image E, which is photographed by camera F.}
\label{setup}
\end{center}
\end{figure}

To demonstrate that our numerically observed phenomenon occurs also in physical sheets, a crude experiment was performed. First, we made a $d$-cone by pushing a reflective plastic sheet into a circular container of radius 9.1cm. The sheet is a an ordinary 22cm $\times$ 28cm projection transparency used in presentations. Its thickness is about 0.1mm. We measured the curvature profile of this sheet by observing the images of standard objects reflected in the sheet. The magnification of each object allows us to infer the curvature at the corresponding position on the $d$-cone. The experimental setup is sketched in FIG.~\ref{setup}. The standard objects are identical black circles arranged in a row, printed on a second transparency. The row of circles is oriented such that its images on the surface of the $d$-cone lie along a radial line in the unbuckled region. The images of the circles are turned into ellipses that are progressively wider in the azimuthal direction as they approach the vertex, since the azimuthal curvature is inversely proportional to the distance to the vertex. In the noncontacting region, the images all have the same height in the radial direction, since radial curvature is zero there. However, the image on the edge is different. It is being contracted in the radial direction due to the induced radial curvature there. If the radial curvature is of the same magnitude as azimuthal curvature at the edge, the contraction in the radial direction and the expansion in the azimuthal direction have to be by the same factor. Let $w$ be the azimuthal size and $l$ be the radial size of the image, then the $w/l$ ratio for the image at the edge has to be the square of that ratio for the image just outside or inside the edge. 

\begin{figure}[!hbt]
\begin{center}
\includegraphics[width=0.65\textwidth, bb=0pt 0pt 411pt 428pt]{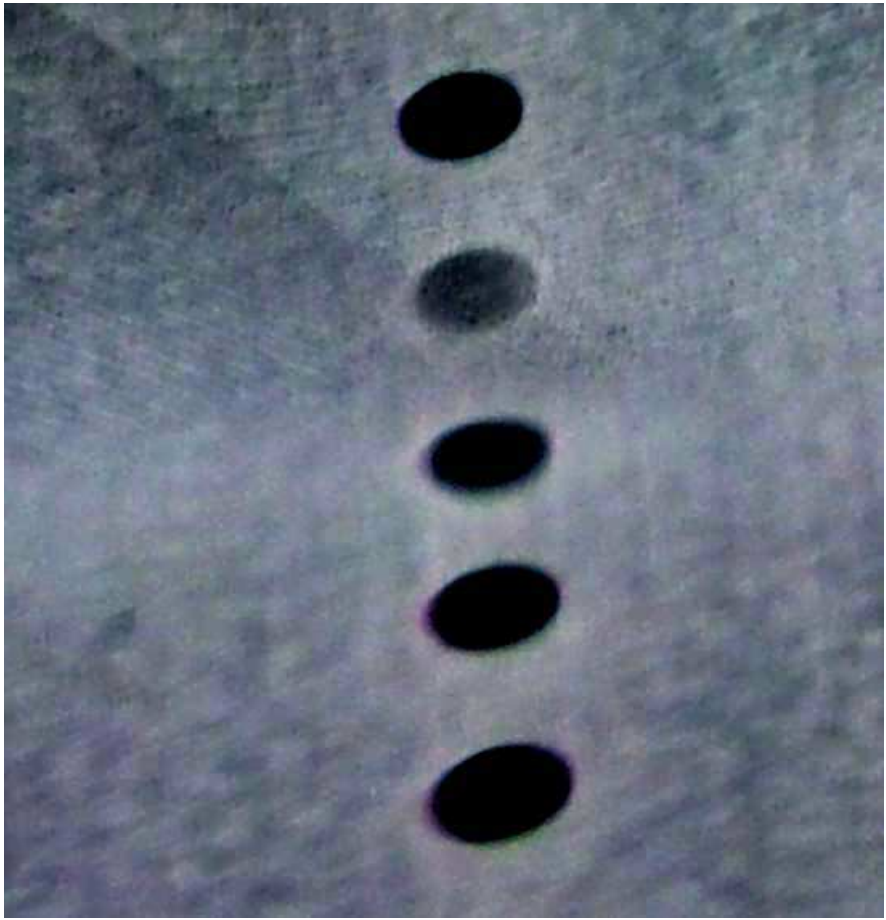}
\caption{(Color online) Images of circles as reflected from the surface of a $d$-cone with $\epsilon \approx 0.25$. The lower images are closer to the tip. The third image lies on the edge, which is perpendicular to the line of images.}
\label{pic1}
\end{center}
\end{figure}

\begin{figure}[!hbt]
\begin{center}
\includegraphics[width=0.65\textwidth, bb=0pt 0pt 506pt 519pt]{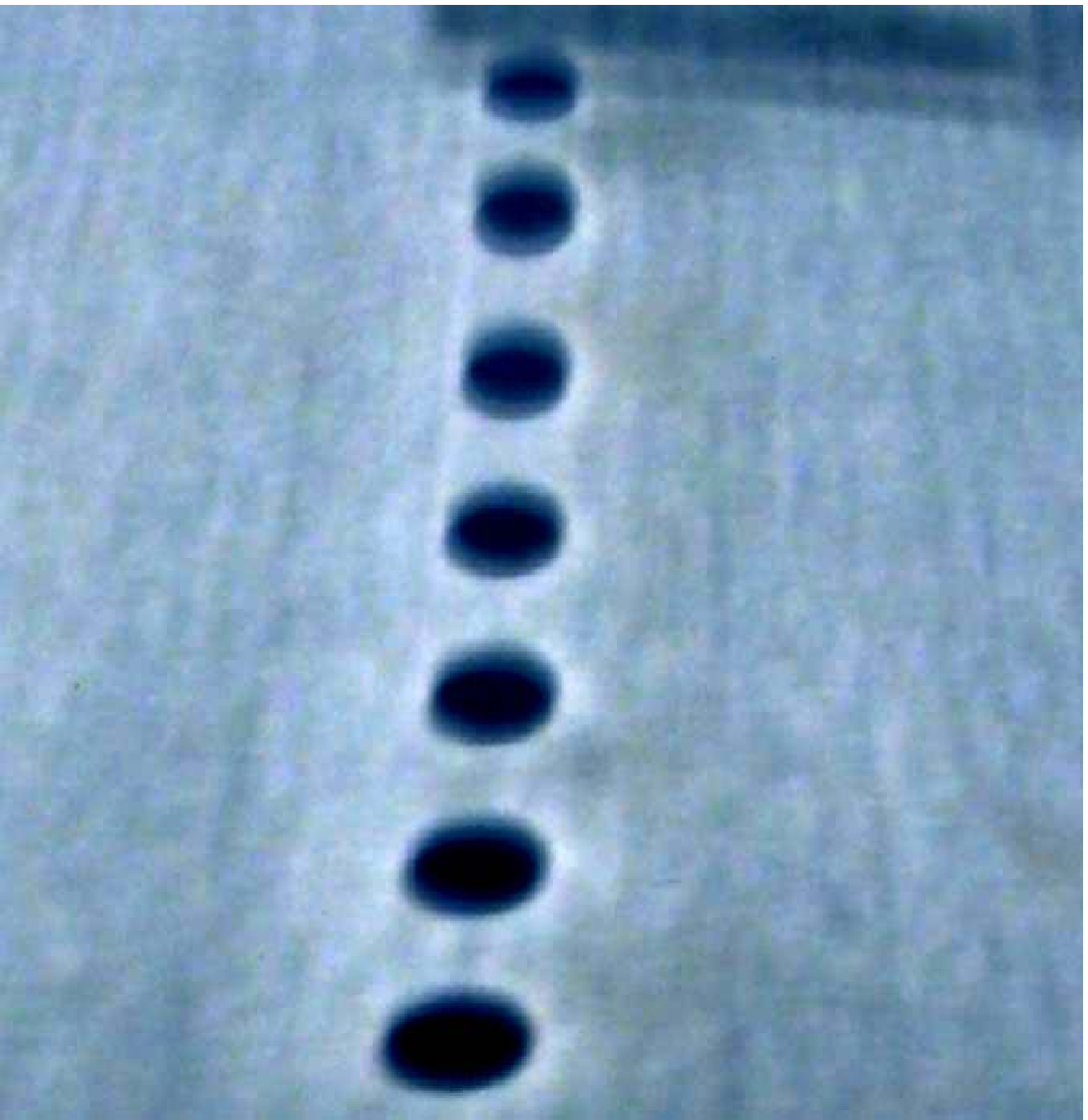}
\caption{(Color online) Images of circles as reflected from the surface of a $d$-cone with $\epsilon \approx 0.15$. The lower images are closer to the tip. The first image from the top lies on the edge, which is perpendicular to the line of images. The images are a little out of focus. The sizes are measured based on the black core region, excluding the blurry fringes.}
\label{pic2}
\end{center}
\end{figure}

We used a digital camera to take pictures of the images near the edge. FIG.~\ref{pic1} is one of the pictures taken for $\epsilon \approx 0.25$. It shows images of 5 circles that lie along a radial direction. The lower images are closer to the tip. The central spot is the one that lies on the edge. The values of azimuthal and radial sizes of the images as well as the their ratios are listed in table I, with the corresponding ratios that show comparison highlighted. From the table, we can see that the expected relationship between $w/l$ ratio of the image that lies on the edge and that of the adjacent image is favored by the data, indicating that radial curvature is indeed of the same magnitude as azimuthal curvature. As another check, FIG.~\ref{pic2} is the picture taken for $\epsilon \approx 0.15$. It shows images of 7 circles. The images are a little out of focus, and we measure their sizes according to the black core region, excluding the blurry fringes. The top spot lies on the edge. Table II gives the values of their sizes and corresponding ratios. We observe that these data are also in agreement with our numerical observation.  

\begin{center}
\begin{table}
\begin{tabular}{ccccccccc}
\hline
&& Azimuthal Size & & Radial Size & & Ratio &&  \\
Spot && $w$ && $l$ && $w/l$  && $(w/l)^2$ \\
\hline
1 && 20 && 15 && 1.333 && \\
2 && 20.5 && 15 && 1.367 && {\bf 1.868} \\
3 && 20.5 && 11 && {\bf 1.864} && \\
4 && 21 && 15 && 1.400 && {\bf 1.960} \\
5 && 22 && 15.5 && 1.419 && \\
\hline
\end{tabular}
\caption{The sizes of images in FIG.~\ref{pic1}, from top to bottom, in arbitrary but fixed units. $w$ is the azimuthal size (width) of the image; $l$ is the radial size (height). The third spot lies on the edge. The actual size of the top spot is about 1.5mm in width and 1mm in height.}
\end{table}
\end{center}

\begin{center}
\begin{table}
\begin{tabular}{ccccccccc}
\hline
&& Azimuthal Size && Radial Size && Ratio &&  \\
Spot && $w$ && $l$ && $w/l$  && $(w/l)^2$ \\
\hline
1 && 16 && 14 && {\bf 1.143} &&  \\
2 && 17 && 16 && 1.063  && {\bf 1.130} \\
3 && 18 && 16.5 && 1.091 && \\
4 && 19 && 16 && 1.188 &&  \\
5 && 22 && 17 && 1.294 && \\
6 && 24 && 17 && 1.412 && \\
7 && 25 && 17 && 1.471 && \\
\hline
\end{tabular}
\caption{The sizes of images in FIG.~\ref{pic2}, from top to bottom, in arbitrary but fixed units. $w$ is the azimuthal size (width) of the image; $l$ is the radial size (height). The top one lies on the edge. The actual size of the top spot is about 3mm in width and 2mm in height.}
\end{table}
\end{center}

\section{Discussion and Conclusion}
In this paper, we have presented our findings of vanishing mean curvature in the contacting region in a single developable cone. The mean curvature near the supporting edge is observed to vanish within our numerical precision for $d$-cones with various confining radii, thicknesses and deformations. We investigate this phenomenon by studying variants of $d$-cone and identifying the features that are important for the vanishing of mean curvature to happen. It is found that the presence of appropriate radial stress is necessary. We also find that the $d$-cone in a circular container somehow produces the right amount of normal forces that induce just enough radial curvature to cancel azimuthal curvature. By comparing $d$-cones formed with circular and elliptical confining edges, we find that circular symmetry is indispensable for the mean curvature to vanish. The numerically observed phenomenon is verified by a crude experiment of projecting identical circles onto the $d$-cone surface and comparing the corresponding magnification ratios. 

It is worth noting that the cancellation of curvatures does not depend on particular choices of edge potential. In our assessment as long as the choices of potential represent a sharp edge, as is the case for our choices presented in Section II, the induced radial curvature cancels the azimuthal curvature at the edge. Some supporting evidence is as follows. If the radial curvature were induced due to the particular form of the edge potential, that radial curvature would be independent of the deflection of the sheet. However, this contradicts with our observations that induced radial curvature does depend on the deflection of the sheet.

The vanishing of mean curvature is precisely the property of a ``minimal surface'', i.e. a surface of minimal surface area for given boundary conditions. Liquidlike membranes, such as a soap film \cite{soapfilm}, follow the shape of a minimal surface in equilibrium and thus have zero mean curvature everywhere. This phenomenon is purely geometric and it reminds one of the Gauss-Bonnet theorem, which constrains the average Gaussian curvature in a surface \cite{struik}. Our results for elliptical container suggest that it is some average mean curvature along the rim that vanishes. Understanding radial curvature $C_{rr}$ requires more than \textit{Elastica} equation, since $C_{rr}$ necessarily creates stretching not allowed in the \textit{Elastica} approach \cite{maha-new}. One approach to understanding the $C_{rr}$ is to assume the contact force inferred from the \textit{Elastica} approach and use the force von \karman~equation to infer $C_{rr}$. This unexpected geometrical regularity in force thin sheets is important to understand, since it surely influences a broad range of situations extending far beyond those studied here. We shall explore this phenomenon further in future publications.  

\begin{acknowledgments}
The authors would like to thank Enrique Cerda, Leo Kadanoff and Paul Goldbart for enlightening discussions. This work was supported in part by the National Science Foundation its MRSEC Program under Award Number DMR-0213745.
\end{acknowledgments}

\end{document}